\documentclass[a4paper]{article}

\usepackage{amssymb}
\usepackage{latexsym}

\newcommand{\secref}[1]{\S~\ref{#1}}
\newcommand{\eqref}[1]{Eq.~(\ref{#1})}
\newcommand{\D}{\mathrm{d}}
\newcommand{\Punto}{\cdot}
\newcommand{\Ndoppia}{{I\!\!N}}
\newcommand{\Rdoppia}{{I\!\!R}}

\newtheorem{proposition}{Proposition}

\newtheorem{lemma}{Lemma}
\newenvironment{proof}%
{\noindent\textit{Proof}.}%
{\hfill$\Box$\\}

\begin{document}

\title{On deformation of Poisson manifolds of hydrodynamic type
\thanks{Work sponsored by the Italian Ministry of Research under the
project 40\%: \textit{Geometry
of Integrable Systems}}}
\author{Luca Degiovanni$^1$
\and Franco Magri$^2$
\and Vincenzo Sciacca$^3$}

\date{}

\maketitle

\begin{enumerate}
\footnotesize
\item Dottorato in Matematica, University of Torino\\
via C. Alberto 10, 10123 Torino\\
\emph{email}: degio@dm.unito.it
\item Department of Mathematics and Applications, University of Milano
Bicocca\\
via degli Arcimboldi 8, 20126 Milano\\
\emph{email}: magri@matapp.unimib.it
\item Dottorato in Matematica, University of Palermo\\
via Archirafi 34, 90123 Palermo\\
\emph{email}: sciacca@dipmat.math.unipa.it
\end{enumerate}

\begin{abstract}
We study a class of deformations of infinite-dimensional Poisson manifolds
of hydrodynamic type
which are of interest in the theory of Frobenius manifolds. We prove two
results. First, we show
that the second cohomology group of these manifolds, in the
Poisson-Lichnerowicz cohomology, is
``essentially'' trivial. Then, we prove a conjecture of B. Dubrovin about
the triviality of
homogeneous formal deformations of the above manifolds.
\end{abstract}

\section{Dubrovin's conjecture}\label{sec:1}

In this paper we solve a problem
proposed by B. Dubrovin in the framework of the theory of Frobenius
manifolds \cite{DB1}. It
concerns the deformations of Poisson tensors of hydrodynamic type. The
challenge is to show that a
large class of these deformations are trivial.

In an epitomized form the problem can be stated as follows.
Let $M$ be a Poisson
manifold endowed with a
Poisson bivector $P_0$ fulfilling the Jacobi condition
\begin{displaymath}
[P_0,P_0] = 0
\end{displaymath}
with respect to the Schouten bracket on the algebra of multivector
fields on $M$. A
deformation of $P_0$ is a formal series
\begin{displaymath}
P_\epsilon = P_0 + \epsilon P_1 + \epsilon^2 P_2 + \cdots
\end{displaymath}
in the space of bivector fields on $M$ satisfying the Jacobi condition
\begin{equation}\label{chiusura}
[P_\epsilon,P_\epsilon] = 0
\end{equation}
for any value of the parameter $\epsilon$.
The deformation is trivial if there exists a formal diffeomorphism
 $\phi_\epsilon : M \rightarrow M$, admitting the Taylor expansion
\begin{displaymath}
\phi_\epsilon = \phi_0 + \epsilon \phi_1 + \epsilon^2 \phi_2 + \cdots,
\end{displaymath}
which pulls back $P_\epsilon$ to $P_0$:
\begin{displaymath}
P_\epsilon = {\phi_\epsilon}_\ast (P_0).
\end{displaymath}
Assume that the class of deformations $P_\epsilon$ and of diffeomorphisms
$\phi_\epsilon$ is
restricted by a set of additional conditions to be described below. The
demand is to prove that
every allowed deformation is trivial, and to provide an explicit procedure
to construct the
trivializing map $\phi_\epsilon$ in the class of allowed transformations.

In the  concrete form suggested by Dubrovin, the manifold $M$ is very
simple but the class of allowed deformations is rather large. That is
the source of difficulty of the problem. Indeed, the manifold $M$ is
the space  of $C^\infty$-maps $u^a(x)$ from $S^1$ into $\Rdoppia^n$,
and the bivector $P_{0}$ is of hydrodynamic type \cite{DB2}. By using the
so-called ``flat coordinates'' $u^{a}$ in $\Rdoppia^n$, it can be written
in the simple form
\begin{displaymath}
P_0 = g^{ab} {\frac{\D }{\D x}}
\end{displaymath}
where the coefficients $g^{ab}$ are the entries of a constant, regular,
symmetric $n \times n$
matrix (not necessarily positive definite). The allowed deformations
$P_\epsilon$ are formal series
of matrix-valued differential operators. The coefficient $P_k$ has degree
$k+1$, and is written in
the form
\begin{displaymath}
P_k = A_0(u) {\frac{\D^{k+1}}{\D x^{k+1}}} + A_1(u;u_x){\frac{\D^{k}}{\D
x^{k}}} + \cdots +
A_{k+1}(u;u_x,\ldots,u_{k+1}).
\end{displaymath}
The entries of the matrix coefficient $A_l$ are assumed to be
homogeneous polynomials of degree $l$ in the derivatives of the field
functions $u^a(x)$. The
degree of a polynomial is computed by attributing degree zero to the field
functions, degree one to
their first derivatives, degree two to the second derivatives, and so on. By
this requirement the form of the operator $P_k$ is fixed up to the
choice of a increasing number of arbitrary functions of the coordinates
$u^a$. These
functions, finally, must be chosen so to guarantee that the operator $P_k$
is skewsymmetric
\begin{equation}\label{antisim}
{P_k}^{*}= -P_k
\end{equation}
and that the Jacobi condition (\ref{chiusura}) is satisfied at the order
$k$. This means that the
first $k$ operators $P_l$ must be chosen so to verify the $k$ conditions
\begin{displaymath}
\sum_{i+j=l}[P_{i},P_{j}] = 0 \qquad l=1, \ldots , k
\end{displaymath}
or, explicitly,
\begin{equation}\label{chiusura2}
\begin{array}{l}
2[P_0,P_1] = 0 \\
2[P_0,P_2] + [P_1,P_1] = 0 \\
2[P_0,P_3] + 2[P_1,P_2] = 0 \\
\quad\cdots
\end{array}
\end{equation}
and so on. The conjecture of Dubrovin is that all these homogeneous
deformations are trivial, and
that the trivializing map is homogeneous as well.

To better understand the problem, let us consider the scalar-valued case.
According to the
rules of the game the first three coefficients of the
deformations $P_\epsilon$ have the form
\begin{eqnarray*}
P_0 &=& \frac{\D}{\D x} \\
P_1  &=& A(u){\frac{\D^2}{\D x^2}} + B(u)u_x{\frac{\D}{\D x}} \\
&& +(C(u)u_{xx}+D(u){u_x}^2) \\
P_2 &=& E(u){\frac{\D^3}{\D x^3}} + F(u)u_x{\frac{\D^2}{\D x^2}} +
(G(u)u_{xx}+H(u){u_x}^2){\frac{\D}{\D x}} \\
&& + (L(u)u_{xxx}+M(u)u_{xx}u_x+N(u){u_x}^3).
\end{eqnarray*}
They depend on eleven arbitrary functions of the coordinate $u$. By
imposing the skewsymmetry
condition (\ref{antisim}) this number falls to four. Indeed we get the
seven differential
constraints
\begin{displaymath}
\begin{array}{l}
A=0\,,\; 2C=B\,,\; 2D=B^{'} \\
2F=3E^{'}\,,\; 4L=2G-E^{'}\,,\; 4N=2H^{'}-E^{'''} \\
4M=2G^{'}+4H-3E^{''}.
\end{array}
\end{displaymath}
The remaining four functions are constrained by the Jacobi condition.
To work out this condition we use the operator form of the Schouten bracket
of two skewsymmetric
operators $P$ and $Q$. We need the following notations. We denote by
$Q_{u}\alpha$ the value of the
differential operator $Q_{u}$ on the argument $\alpha$, and by
\begin{displaymath}
Q_{u}^{'}(\alpha ;\dot{u})={\frac{\D }{\D s}}Q_{u+s\dot{u}}\alpha |_{s=0}
\end{displaymath}
its derivative along the vector field $\dot{u}$. The adjoint of this
derivative with respect to
$\dot{u}$ is denoted by ${Q_{u}^{'}}^{*}(\alpha;\beta)$. It is defined by
\begin{displaymath}
\langle Q_{u}^{'}(\alpha ;\dot{u}),\beta \rangle=\langle
\dot{u},{Q_{u}^{'}}^{*}(\alpha;\beta) \rangle ,
\end{displaymath}
where the pairing between vector fields and $1$-forms is defined, as usual, by
\begin{displaymath}
\langle \dot{u},\beta \rangle =\int_{S^{1}}\dot{u}(x) \beta(x) \D x .
\end{displaymath}
(Of course, in the vector-valued case we have to sum over the different
components).
Then the Schouten bracket is given by \cite{MM}:
\begin{eqnarray*}
2[P,Q](\alpha,\beta) &=& P_{u}^{'}(\alpha;Q_{u}\beta) -
P_{u}^{'}(\beta;Q_{u}\alpha) - Q_{u}\Punto{P_{u}^{'}}^{*}(\alpha;\beta)\\
&& + Q_{u}^{'}(\alpha;P_{u}\beta)-Q_{u}^{'}(\beta;P_{u}\alpha) -
P_{u}\Punto{Q_{u}^{'}}^{*}(\alpha;\beta).
\end{eqnarray*}
In our example, the bivector $P_{0}$ is constant. Therefore, the first two
Jacobi conditions
(\ref{chiusura2}) take the simple form:
\begin{displaymath}
{P_{1}}^{'}_u(\alpha;P_{0}\beta) - {P_1}^{'}_u(\beta;P_0\alpha) -
P_0\Punto{{P_1}^{'}_u}^{*}(\alpha;\beta)=0
\end{displaymath}
and
\begin{eqnarray*}
&&{P_2}^{'}_u(\alpha;P_0\beta) - {P_2}^{'}_{u}(\beta;P_0\alpha) -
P_0\Punto{{P_2}^{'}_{u}}^{*}(\alpha;\beta)\\
&&+{P_1}^{'}_{u}(\alpha;{P_1}_{u}\beta) -
{P_1}^{'}_{u}(\beta;{P_1}_{u}\alpha) -
{P_1}_{u}\Punto{{P_1}^{'}_{u}}^{*}(\alpha;\beta)=0
\end{eqnarray*}
respectively. By expanding these operator conditions, we obtain two further
relations
\begin{eqnarray*}
B&=&0 \\
4H&=&2G^{'}+E^{''}
\end{eqnarray*}
among the eleven functions $(A(u), \ldots ,N(u))$. Solving them and the
previous ones we obtain
that the first coefficients of $P_\epsilon$ are:
\begin{eqnarray}\label{piuno}
P_1 &=& 0 \\ \label{pidue}
P_2 &=& E(u){\frac{\D^3}{\D x^3}} +{\frac{3}{2}} E^{'}(u)u_x{\frac{\D^2}{\D
x^2}} +
G(u)u_{xx}{\frac{\D }{\D x}}\\ \nonumber
&& +\frac{1}{2}(G^{'}(u) + {\frac{1}{2}}E^{''}(u)){u_x}^2{\frac{\D}{\D x}} +
\frac{1}{2}(G(u) - \frac{1}{2}E^{'}(u))u_{xxx} \\
&& +(G^{'}(u) - {\frac{1}{2}}E^{''}(u))u_{xx}u_{x} +
\frac{1}{4}(G^{''}(u) - {\frac{1}{2}}E^{'''}(u)){u_x}^3. \nonumber
\end{eqnarray}
Up to the second order in $\epsilon$, this is the most general homogeneous
deformation in the
scalar case.

To check Dubrovin's conjecture to the second-order in $\epsilon$, it is
enough to consider the
homogeneous map
\begin{displaymath}
\phi_{\epsilon}(u)=u+{\epsilon}R(u)u_{x}+{\epsilon}^{2}(S(u)u_{xx}+T(u)u_{x}^{2}
)+\dots
\end{displaymath}
and to use the operator form \cite{MM}
\begin{equation}\label{transf}
P_\epsilon={{\phi}^{'}_{\epsilon}}_{u}{\Punto}P_0\Punto{{\phi}^{'}_{\epsilon}}_{
u}^{*}
\end{equation}
of the transformation law for bivectors. As before,
${{\phi}^{'}_{\epsilon}}_{u}^{*}$ denotes the
adjoint operator of the Fr\'echet derivative of ${\phi}_{\epsilon}(u).$ By
expanding
\eqref{transf}, we find:
\begin{eqnarray*}
P_{1} &=& 0\\
P_2 &=&(2S(u)-R^{2}(u)){\frac{\D^3}{\D x^3}}
+3(S^{'}(u)-R(u)R^{'}(u))u_x{\frac{\D^2}{\D x^2}}\\
&&+(5S^{'}(u)-4T(u)-R(u)R^{'}(u))u_{xx}{\frac{\D }{\D x}}\\
&&+(3S^{''}(u)-2T^{'}(u)-R(u)R^{''}(u)-{R^{'}(u)}^{2}){u_x}^2{\frac{\D }{\D
x}} \\
&&+2(S^{'}(u)-T(u))u_{xxx}+4(S^{''}(u)-T^{'}(u))u_{xx}u_x\\
&&+(S^{'''}(u)-T^{''}(u)){u_x}^3.
\end{eqnarray*}
By comparison with \eqref{piuno} and \eqref{pidue}, we realize that
Dubrovin's conjecture is true in
the scalar case, up to the second order in $\epsilon$. In fact the choices
\begin{eqnarray}\label{relaz}
R(u)&=& 0 \nonumber\\
S(u)&=& {\frac{1}{2}}E(u) \\
T(u)&=& {\frac{5}{8}}E^{'}(u)-{\frac{1}{4}}G(u) \nonumber
\end{eqnarray}
allow to reconstruct the diffeomorphism $\phi_{\epsilon}$ from the
deformation $P_{\epsilon}$.

The questions now are: what happens at higher order in $\epsilon$, or in
the matrix case? What is
the meaning of the relations (\ref{relaz}) connecting $P_{\epsilon}$ to
$\phi_{\epsilon}$? Due to
the great complexity of the computations, it is clear that any direct
attack is beyond our reach.
We have to devise an alternative approach. Our strategy is to convert the
given problem into a
problem in Poisson-Lichnerowicz cohomology. It is based on two remarks:

\begin{enumerate}
\item Poisson manifold of hydrodynamic type are transversally constant.
\item The second cohomology group in the Poisson-Lichnerowicz
cohomology of these manifolds is ``essentially'' trivial.
\end{enumerate}

The first remark concerns the symplectic foliation associated with the
Poisson bivector $P_0$. In
our example, this foliation is rather regular. All the leaves are affine
hyperplane of codimension
$n$. They are the level sets of $n$ globally defined \emph{Casimir
functions} $C^a$, \mbox{$a=1,2,
\dots ,n$}. Furthermore there exists an abelian group of symplectic
diffeomorphisms which transform
the symplectic leaves among themselves.

The second remark concerns the bivectors $Q$ fulfilling the condition
\begin{displaymath}
[P_{0},Q]=0 .
\end{displaymath}
They must be compared with the bivectors
\begin{displaymath}
Q=L_{X}(P_{0})
\end{displaymath}
which are Lie derivatives of $P_0$ along any vector field $X$ on $M$. The
former are called
$2$-cocycles in the Poisson-Lichnerowicz cohomology defined by $P_0$ on $M$
\cite{LP}. The latter
are called
$2$-coboundaries. Not all cocycles are coboundaries. A first simple
obstruction is the vanishing of
the Poisson bracket of the Casimir functions $C^a$ with respect to $Q$:
\begin{equation}\label{casimir}
\{C^{a}, C^{b}\}_{Q}=0 .
\end{equation}
Further obstructions depend on the topology of the manifold. The main
result of the paper is the
proof, in \secref{sec:3}, that these further obstructions are absent on a
Poisson manifold of
hydrodynamic type. By a combined use of ideas of the theory of
transversally constant Poisson
manifolds (suitably extended to infinite-dimensional manifolds) and of the
operator approach to the
inverse problem of the Calculus of Variations in the style of Volterra
\cite{Vo1} \cite{Vo2}, we show that
every
$2$-cocycle verifying \eqref{casimir} is a $2$-coboundary, and we give  an
explicit formula for
the vector field $X$ (called the potential of $Q$). Several examples of
this result are shown in \secref{sec:4}, where possible applications to the
classifications of bihamiltonian manifolds are also briefly discussed.

Once equipped with this result, the conjecture of Dubrovin can be
proved in a direct and simple way. First we notice that the homogeneous
deformations pass the
obstructions (\ref{casimir}). Then we notice that the Jacobi condition $\left[
P_{\epsilon},P_{\epsilon} \right]=0$ may be replaced by a recursive system
of cohomological
equations. This leads to a simple general representation of the deformation
$P_{\epsilon}$. The
argument goes as follows. Consider the first Jacobi condition
\begin{displaymath}
\left[ P_{0},P_{1} \right]=0.
\end{displaymath}
It is already in a cohomological form. By the main result of
\secref{sec:3}, it follows that there
exists a vector field $X_1$, such that
\begin{displaymath}
P_{1}=L_{X_1}(P_{0}).
\end{displaymath}
By inserting this information into the second Jacobi equation
\begin{displaymath}
2\left[ P_{0},P_{2} \right]+\left[ P_{1},P_{1} \right]=0
\end{displaymath}
we get a new cohomological equation
\begin{displaymath}
\left[ P_{0},P_{2}-{\frac{1}{2}}L^{2}_{X_1}(P_{0}) \right]=0 .
\end{displaymath}
hence there exists a second vector field $X_2$ such that
\begin{displaymath}
P_{2}=L_{X_2}(P_{0})+{\frac{1}{2}}L^{2}_{X_1}(P_{0}) .
\end{displaymath}
By induction, one proves the existence of a sequence of vector fields
$\{X_{k}\}_{k
\in{\Ndoppia}}$ such that all the coefficients $P_k$ of the deformation
$P_\epsilon$ admits the
representation
\begin{equation}\label{Lieformula}
P_{k}=\sum_{j_{1}+2j_{2}+\dots +kj_{k}=k}
\left(\frac{L^{j_{k}}_{X_{k}}}{j_{k}!}\frac{L^{j_{k-1}}_{X_{k-1}}}{j_{k-1}!} 
\cdots
\frac{L^{j_{1}}_{X_{1}}}{j_{1}!}\right) (P_{0}).
\end{equation}
This result gives a complete control of the deformations of Poisson
brackets of hydrodynamic type. In particular it allows to give the
following simple proof of the Dubrovin's conjecture. Consider
separately the different flows ${\phi}^{(k)}_{t_{k}}$ associated with the
vector fields $X_k$. Give
them a different weight by setting
\begin{displaymath}
t_{k}={\epsilon}^{k} ,
\end{displaymath}
and make the ordered product of these flows by multiplying each flow by the
subsequent one on the
left. The result is the one-parameter family of diffeomorphism
\begin{displaymath}
{\phi}_{\epsilon}=\prod_{k\ge 1}{\phi}^{(k)}_{{\epsilon}^k}.
\end{displaymath}
It provides the solution we were looking for. Indeed, according to the
theory of ``Lie
transform'', \eqref{Lieformula} are equivalent to the transformation law
\begin{displaymath}
P_{\epsilon}={{\phi}_{\epsilon}}_{*}(P_{0})
\end{displaymath}
as required. We believe that the strategy sketched above is of interest in
itself, and that it can
be profitably used in more general context. In our opinion it can provide,
for instance, new
insights on the problem of classification of bihamiltonian manifolds
associated with soliton
equations.

\section{Transversally constant Poisson manifolds}\label{sec:2}

In this section we collect the few ideas of the theory of Poisson
manifolds which are used later on to solve Dubrovin's
conjecture. Our interest is mainly centered around the difference
between $2$-cocycles and $2$-coboundaries on a regular transversally
constant Poisson manifold.

We recall that a finite-dimensional Poisson manifold $(M,P)$ is
\emph{regular} if the symplectic foliation defined by the Poisson
bivector $P$ has constant rank. Let $k$ denote the corank of the
foliation. It follows that, around any point of the manifold, there
exist $k$ functions $C^a$, $a=1,2,\dots ,k$, which are independent and
constant on each symplectic leaf. They are called Casimir
functions. Their differentials $\D C^a$ span the kernel of the bivector
$P$. We also recall that the Poisson manifold is called
\emph{transversally constant} \cite{Va} if there exist $k$ vector fields
$Z_a$ which are transversal to the symplectic leaves and are
symmetries of $P$:
\begin{displaymath}
L_{Z^a}(P)=0.
\end{displaymath}
Without loss of generality, one can assume that these vector fields
satisfy the normalization conditions
\begin{displaymath}
Z_{a}(C^b)={\delta}^{b}_{a}
\end{displaymath}
with respect to the chosen family of Casimir functions.

The local structure of a transversally constant Poisson manifold is
quite simple: essentially is the product of a symplectic leaf and of
the abelian group generated by the vector fields $Z_a$. In particular,
the tangent space at any point $m$ can be split into the direct sum
\begin{displaymath}
T_{m}M=H_{m}\oplus V_{m}
\end{displaymath}
of an ``horizontal space'' $H_m$ (the tangent space of the symplectic
leaf) and of a ``vertical space'' $V_m$, spanned by the vector fields
$Z_a$. This splitting induces a corresponding decomposition of the
dual space and, hence, of any tensor field on $M$. For a bivector $Q$
the basic elements are the vector fields
\begin{displaymath}
X^{a}=Q\D C^a
\end{displaymath}
and the horizontal bivector
\begin{displaymath}
Q_{H}={\pi}_{H}\circ Q\circ{{\pi}_{H}}^{*}
\end{displaymath}
where $\pi_{H}$ denotes, as usual, the canonical projection on $H$
along $V$. A simple computation gives
\begin{equation}\label{definizione1}
Q_{H}=Q+X^{a}\wedge Z_{a}+{\frac{1}{2}}X^{a}(C^b)Z_{a}\wedge Z_{b}.
\end{equation}
They already contain the clue of the distinction between $2$-cocycle
and $2$-coboundaries.

\begin{lemma}\label{lemma1}
If $Q$ is a cocycle the vector fields $X^a$ are symmetries of $P$ and
$Q_H$ is a cocycle. If $Q$ is a coboundary the vector fields $X^a$ are
Hamiltonian and $Q_H$ is a coboundary.
\end{lemma}
\begin{proof}
If $Q$ is a cocycle we have
\begin{displaymath}
L_{Q\D F}(P)+L_{P\D F}(Q)=0
\end{displaymath}
for any function $F$. For $F=C^a$, this equation shows that $X^a$ is a
symmetry of $P.$ Hence
\begin{displaymath}
\left[ P, X^{a}\wedge Z_{a}+{\frac{1}{2}}X^{a}(C^b)Z_{a}\wedge Z_{b}
\right]=0
\end{displaymath}
since both $X^a$ and $Z_a$ are symmetries of $P$ and $X^{a}(C^b)$ is a
Casimir function. This show that
\begin{displaymath}
\left[ P,Q_H \right]=0
\end{displaymath}
as claimed.

If $Q=L_{X}(P)$ is a coboundary, we find
\begin{displaymath}
Q\D C^a = L_{X}(P)\D C^a =L_{X}(P\D C^a)- P\D X(C^a)=-P\D X(C^a)
\end{displaymath}
showing that
$$Q\D C^a=P\D H^a$$
with
$$H^a =-X(C^a).$$
Therefore we find
$$X^{a}\wedge Z_{a}+{\frac{1}{2}}X^{a}(C^b)Z_{a}\wedge
Z_{b}=L_{Z}(P)$$
with
$$Z=-H^{a}Z_{a}.$$
This proves the second part of the Lemma.
\end{proof}

The previous remark alone is sufficient for the later
applications. However, in view of adapting the result to the case of
infinite-dimensional Poisson manifolds of hydrodynamic type, it is
better to restate it in a different form. The idea is to trade
multivectors for forms. To this end, we first split the vector fields
$X^a$ into horizontal and vertical parts. Then, the components of the
vertical parts are used to define the matrix
\begin{displaymath}
\{C^{a},C^{b}\}_{Q}:=X^{a}(C^b).
\end{displaymath}
The horizontal parts $X^{a}_{H}$ are, instead, used to define $k$
$1$-forms $\theta^a$ living on the symplectic leaves. They are given
by
\begin{equation}\label{definizione2}
{\theta}^{a}(X_F)=X^{a}_{H}(F)
\end{equation}
where $X_F =P\D F$ is the Hamiltonian vector field associated with the
function $F$. Similarly, the horizontal bivector $Q_H$ is traded for a
$2$-form $\omega$, living on the symplectic leaves, according to
\begin{displaymath}
{\omega}(X_F ,X_G )=Q_{H}(\D F,\D G).
\end{displaymath}
The outcome is that any bivector $Q$ on a regular transversally
constant Poisson manifold $M$ is characterized by three elements:

\begin{enumerate}
\item the functions $\{ C^{a},C^{b} \}_{Q}$
\item the $1$-forms $\theta^a$
\item the $2$-form $\omega$.
\end{enumerate}

As a simple restatement of the previous Lemma, we obtain the following
result.

\begin{lemma}
If $Q$ is a cocycle $\{ C^{a},C^{b} \}_{Q}$ is a Casimir
function, and the forms $\theta^a$ and $\omega$ are closed. If $Q$ is
a coboundary the functions $\{ C^{a},C^{b} \}_{Q}$ vanish,
and the forms $\theta^a$ and $\omega$ are exact.
\end{lemma}

We do not give the proof of this result, that can be found in
\cite{Va}. Instead, for further convenience, we show its converse in the
following form.

\begin{lemma}
If the functions $\{ C^{a},C^{b} \}_{Q}$ vanish,
\begin{equation}\label{esattezza1}
\{ C^{a},C^{b} \}_{Q}=0,
\end{equation}
and the forms $\theta^a$ and $\omega$ are exact,
\begin{equation}\label{esattezza2}
\theta^{a}=\D H^{a}
\end{equation}
\begin{equation}\label{esattezza3}
\omega=d\theta,
\end{equation}
the bivector $Q$ is a coboundary. Its potential $X$ is given by
\begin{equation}\label{potenziale}
X=-H^{a}Z_{a}+P\theta.
\end{equation}
\end{lemma}
\begin{proof}
The first assumption (\ref{esattezza1}) entails that the vector fields
$X^a$ are tangent to the symplectic leaves. Hence $X^a
=X^{a}_{H}.$ Thus the definition (\ref{definizione2}) and the second
assumption (\ref{esattezza2}) leads to
\begin{displaymath}
X^{a}=P\D H^{a}.
\end{displaymath}
Set $Z=-H^{a}Z_{a}.$ As in the proof of Lemma \ref{lemma1} we get
\begin{equation}\label{scomposizione1}
Q=Q_{H}+L_{Z}(P).
\end{equation}
Finally, we notice that the third assumption (\ref{esattezza3})
entails
$$Q_{H}(\D F,\D G)={\omega}(X_{F},X_{G})=d\theta(X_{F},X_{G})
=L_{P\theta}(P)(\D F,\D G).$$
Hence the previous equation becomes
$$Q=L_{P\theta}(P)+L_{Z}(P)=L_{X}(P)$$
as claimed.
\end{proof}

A difficulty is readily met in trying to extend the previous result to
infinite-dimensional
manifolds. It is connected to the definition (\ref{definizione1}) of the
bivector $Q_H$ where the
operation of exterior product is used. We have found difficult to extend
this formula in the
infinite-dimensional setting where vector fields and bivectors are
represented by differential
operators. To circumvent this difficulty, we can follow a two-steps
procedure, where the vector
fields $X^a$ come first, and only later the bivector $Q_H$ is introduced as
the complementary part
of
$L_{Z}(P)$ in the splitting (\ref{scomposizione1}) of $Q$. This detour
leads to an ``eight steps algorithm'' to check if a given bivector $Q$
on a transversally constant Poisson manifold is a coboundary. They are:
\begin{enumerate}
\item Check that the functions $\{ C^{a},C^{b} \}_{Q}$ vanish.
\item Check that the vector fields $Q\D C^a$ are Hamiltonian:
$Q\D C^{a}=P\D H^{a}$.
\item Introduce the transversal vector field $Z=-H^{a}Z_{a}$.
\item Compute the Lie derivatives of $P$ along $Z$.
\item Define the horizontal bivector $Q_H$ according to:
$Q_{H}=Q-L_{Z}(P)$.
\item Introduce the $2$-form $\omega$ by factorizing $Q_H$ according to:
 $Q_{H}=P\circ \omega\circ P$.
\item Check that this form is exact on the symplectic leaves.
\item Compute its potential $\theta.$
\end{enumerate}
At the end of this long chain of tests, one can affirm that $Q$
is a coboundary and construct its potential $X$ according to
\eqref{potenziale}. In the next section we shall display this procedure for
manifolds of hydrodynamic type.

\section{Poisson manifolds of hydrodynamic type}\label{sec:3}

Let now
$$P=P_0=g^{ab}\frac{\D }{\D x}.$$
We notice that this bivector admits $k$ globally defined Casimir functions
$$C^{a}(u)=\int_{0}^{1}u^{a}(x)\D x.$$
Therefore its symplectic leaves are affine hyperplanes and
the manifold is regular. We also notice that the vector fields
\begin{eqnarray*}
Z_{a}: && \dot{u}^b=\delta_{a}^{b}
\end{eqnarray*}
are globally defined transversal symmetries. Hence, the manifold is
transversally constant as well.

On this manifold we consider the class of bivectors $Q$ which are
represented by matrix-valued differential operators
\begin{equation}\label{bivettoreQ}
Q={\sum}_{k{\geq 0}}A_{k}(u,u_{x},{\dots}){\frac{\D^k}{\D x^k}}
\end{equation}
and which satisfy the simple condition
\begin{equation}\label{cond1}
\{C^{a},C^{b}\}_{Q}=0.
\end{equation}
We stress that no homogeneity conditions are imposed on $Q.$ So the
present class of bivectors is bigger than that considered in
Dubrovin's conjecture. We shall prove

\begin{proposition}\label{prop:1}
Each cocycle $Q$ in this class is a coboundary.
\end{proposition}

To appreciate the strength of this result, let us consider the case of a
single loop function
$u(x)$. Condition (\ref{cond1}) is automatically verified in this case,
since there is only one
Casimir function, and therefore we can conclude that \emph{every}
scalar-valued cocycle is a
coboundary. This result is far from being trivial. Let us check this claim
for the simple cocycle
$P_2$ considered in \secref{sec:1}. We have to exhibit a vector field
$$\dot{u}=X(u,u_{x},u_{xx})$$
such that
$$-P_2=-L_X(P_0)=X^{'}_{u}{\Punto}P_0+P_0\Punto{X^{'}_{u}}^{*}$$
where ${X^{'}_{u}}^{*}$ is the formal adjoint of the Fr\'echet
derivative $X^{'}_{u}$ of the operator defining the vector field $X.$ A
reasonable guess is to look for a homogeneous vector
field
$$X=a(u)u_{xx}+b(u)u_{x}^{2}.$$
Since
$$X^{'}_{u}=a(u){\frac{\D ^2}{\D x^2}}+2b(u)u_{x}{\frac{\D }{\D
x}}+a^{'}(u)u_{xx}+b^{'}(u)u^{2}_{x},$$
a simple computation leads to
\begin{eqnarray*}
-L_{X}(P_0)&=&2a(u){\frac{\D ^3}{\D x^3}}+3a^{'}(u)u_{x}{\frac{\D^2}{\D x^2}}+
(5a^{'}(u)-4b(u))u_{xx}{\frac{\D }{\D x}}\\
&& +(3a^{''}(u)-2b^{'}(u))u^{2}_{x}{\frac{\D }{\D x}} +
2(a^{'}(u)-b(u))u_xxx \\
&& +4(a^{''}(u)-b^{'}(u))u_{x}u_{xx}+(a^{'''}(u)-b^{''}(u))u^{3}_{x}.
\end{eqnarray*}
The problem is solved by noticing that the relations
$$a(u)=-{\frac{1}{2}}E(u)$$
$$b(u)={\frac{1}{4}}G(u)-{\frac{5}{8}}E^{'}(u)$$
allow to identify the operator $L_{X}(P)$ with $P_{2},$ for \emph{any} choice
of the function $E(u)$ and $G(u)$ (see \eqref{pidue}), as claimed in
Proposition~\ref{prop:1}.\\

In this section we shall show that the above relations are simply an  instance
of the general formula (\ref{potenziale}), defining the potential $X$
of any coboundary of a transversally constant Poisson manifold. The main
difficulty is to identify the geometrical objects (the vector
fields $Q\D C^a,$ the $1$-forms $\theta^a,$ and the $2$-form $\omega$)
to be associated with each bivector of the form (\ref{bivettoreQ}). To
this end, it is useful to split the operator $Q$ into the sum of
three operators. The first operator has degree zero. Therefore it is
simply a skewsymmetric matrix $E$, whose entries are functions of the
loops $u^{a}(x)$ and of their derivatives. The second operator has
order one. It is written as the anticommutator
$S\Punto{\frac{\D}{\D x}}+{\frac{\D}{\D x}}\Punto S$ of a symmetric matrix $S$
with ${\frac{\D}{\D x}}$. The third operator, finally, collect all the
higher order terms.

\begin{lemma}
Any bivector $Q$ can be uniquely written in the form:
\begin{equation}\label{rappresentazione}
Q=E+S\Punto{\frac{\D}{\D x}}+{\frac{\D }{\D x}}{\Punto}S+{\frac{\D}{\D
x}}\Punto{\Lambda}\Punto{\frac{\D}{\D x}}
\end{equation}
where $\Lambda$ is the skewsymmetric operator
\begin{displaymath}
{\Lambda}=\sum_{k\geq
0} \left( {\Lambda}_{k}\Punto{\frac{\D^k}{\D{x}^k}}+{\frac{\D^k}{\D
x^k}}\Punto{\Lambda}_{k}
\right).
\end{displaymath}
The coefficients ${\Lambda}_{k}$ of this operator are alternatively
symmetric and skewsymmetric matrices, according to the order of the
derivatives.
\end{lemma}
This lemma is very simple to prove, but it is interesting because each term
in the
splitting (\ref{rappresentazione}) has a geometrical meaning. Roughly
speaking, the first term $E$
controls the brackets
$\{C^{a},C^{b}\}_{Q},$ the second term controls the $1$-forms
$\theta^a,$ and the third term controls the $2$-form $\omega$. By using
this representation
formula we can now work out the ``eight steps algorithm'' stated at the end
of the previous
section.\\

\noindent\textbf{Step 1:} \emph{the vanishing of the functions
$\{C^{a},C^{b}\}_{Q}$.}\\
Since the differentials of the Casimir functions are the constant
matrices
\begin{displaymath}
\frac{{\delta}C^a}{{\delta}u^b}={\delta}^{a}_{b}
\end{displaymath}
we easily find
\begin{displaymath}
\{C^{a},C^{b}\}_{Q}={\int}_{0}^{1}E^{ab}\D x
\end{displaymath}
where $E^{ab}$ is the entry of place $(a,b)$ in the matrix $E$.
Therefore, condition (\ref{cond1}) holds iff there exists a second
skewsymmetric matrix $\mathcal{E}$ such that
$$E={\frac{\D }{\D x}}(\mathcal{E}) .$$
Writing this condition in the commutator form
$$E={\frac{\D }{\D x}}\Punto\mathcal{E}-\mathcal{E}\Punto{\frac{\D }{\D x}}$$
we can easily eliminate $E$ from the representation
(\ref{rappresentazione}) of $Q$. Setting $B=\mathcal{E} +S$ we get
$$Q=B^{t}\Punto{\frac{\D }{\D x}}+{\frac{\D }{\D x}}\Punto B+{\frac{\D }{\D
x}}\Punto{\Lambda}\Punto{\frac{\D }{\D x}}.$$ Finally we replace the
differential operator
$\frac{\D }{\D x}$ by the Poisson bivector
\begin{eqnarray*}
P&=&G\Punto\frac{\D }{\D x}.
\end{eqnarray*}
We then arrive to the following useful second representation theorem.

\begin{lemma}
Each bivector $Q$ for which $\{C^{a},C^{b}\}_{Q}=0$ can be uniquely
represented in the form
\begin{equation}\label{Qnuovissimo}
Q=A^{t}{\Punto}P+P{\Punto}A+P{\Punto}{\Gamma}{\Punto}P
\end{equation}
where $A$ is the $n\times n$ matrix and $\Gamma$ is the skewsymmetric
differential
operator given by:
\begin{eqnarray*}
B&=&G\Punto A \\
\Lambda&=&G\Punto\Gamma\Punto G\,.
\end{eqnarray*}
\end{lemma}

~\\
\noindent\textbf{Step 2:} \emph{the vector fields $Q\D C^a$ are Hamiltonian.}\\
Since \mbox{$\{C^a,C^b\}_Q=0$} we know that the vector fields \mbox{$Q \D
C^a$} are tangent to the
symplectic leaves of $P$. Therefore there exist 1-forms $\theta^a$ such that
\begin{displaymath}
Q \D C^a = P\theta^a.
\end{displaymath}
>From the representation theorem we easily recognize that the 1-form
$\theta^a$ is given by the $a$-th
column of the matrix $A$. So the component $b$ of the 1-form $\theta^a$ is
the entry $A^a_b$ of
place \mbox{$(a,b)$} of the matrix $A$:
\begin{displaymath}
\theta^a_b = A^a_b .
\end{displaymath}
We further know that the vector fields \mbox{$Q \D C^a$} are symmetries of
$P$ by the cocycle
condition. If we work out explicitly the condition
\begin{equation}\label{cociclopiccola}
L_{Q \D C^a}(P)=0
\end{equation}
in the operator formalism we have
\begin{eqnarray*}
-L_{P\theta^a}(P) & = & P\Punto{\theta^{a}}^{'}\Punto P -
P\Punto{{\theta^{a}}^{'}}^{*}\Punto P \\
& = & P\Punto({\theta^{a}}^{'}-{{\theta^{a}}^{'}}^{*})\Punto P = 0.
\end{eqnarray*}
This is the same as writing
\begin{equation}\label{primachiusura}
\frac{\D}{\D
x}{\Punto}({\theta^{a}}^{'}-{{\theta^{a}}^{'}}^{*}){\Punto}\frac{\D}{\D x}
= 0.
\end{equation}
Let us expand the differential operator
\mbox{${\theta^{a}}^{'}-{{\theta^{a}}^{'}}^{*}$} in power
of $\frac{\D}{\D x}$:
\begin{displaymath}
{\theta^{a}}^{'}-{{\theta^{a}}^{'}}^{*} = A_0 + A_1{\Punto}\frac{\D}{\D x}
+\cdots+
A_n{\Punto}\frac{\D}{\D x^n}.
\end{displaymath}
Substituting into the previous equation we obtain
\begin{eqnarray*}
\frac{\D}{\D
x}{\Punto}({\theta^{a}}^{'}-{{\theta^{a}}^{'}}^{*}){\Punto}\frac{\D}{\D x}
&=&A_{0x}{\Punto}\frac{\D}{\D x} + (A_0+A_{1x}){\Punto}\frac{\D^2}{\D x^2}
+\cdots\\
&&+ (A_{n-1}+A_{nx}){\Punto}\frac{\D ^{n+1}}{\D x^{n+1}} +
A_n{\Punto}\frac{\D ^{n+2}}{\D x^{n+2}}
\end{eqnarray*}
showing that the condition (\ref{primachiusura}) can be verified iff
\mbox{${\theta^{a}}^{'}-{{\theta^{a}}^{'}}^{*}=0$}. This means that the
Fr\'echet derivative of the
operator $\theta^a$ is symmetric and, therefore, that this operator is
potential \cite{Pot}. In geometric
language this means that the 1-form $\theta^a$ is closed and therefore
exact, since the
topology of the manifold
$M$ is simple. The potential is the functional
\begin{displaymath}
H^a = \int_0^1 h^a(u,u_x,\ldots)\D x
\end{displaymath}
where, according to \cite{Ton},
\begin{equation}\label{hamiltoniane}
h^a = \int_0^1 A^a_b(\lambda u, \lambda u_x,\ldots) u^b \D\lambda .
\end{equation}
We have thus proved that the vector fields $Q\D C^a$ are Hamiltonian.\\

\noindent\textbf{Step 3:} \emph{the transversal vector field $Z$.}\\
We choose the transversal vector field
\begin{equation}\label{campoZ}
Z = -h^a(u,u_x,\ldots)Z_a
\end{equation}
(sum over repeated index $a$). We notice that by this choice we depart
slightly from the
geometrical scheme. According to the third step of \secref{sec:2} we should
have introduced at this
point the vector field
\begin{displaymath}
\hat{Z} = -\left(\int_0^1 h^a(u,u_x,\ldots)\D x \right) Z_a
\end{displaymath}
whose components are the functionals $H^a$, instead of the associated
densities $h^a$. The change
is permitted since \mbox{$Z(C^a)=\hat{Z}(C^a)$} and so the functions $C^a$
are Casimir functions
also for \mbox{$Q-L_ZP$}. This fact allows to still define a $2$-form
$\omega$ but in general this
is different from the one associated with the previous ``horizontal'' part
of the bivector $Q$. Our choice has the advantage that the vector field $Z$
is local.\\

\noindent\textbf{Step 4:} \emph{the Lie derivative $L_{Z}(P)$.}\\
The next step is to compute the Lie derivative of $P$ along $Z$. In the
operator formalism this is
easily accomplished if we know the Fr\'echet derivative $Z_u'$ of the
vector field $Z$. This
derivative is a matrix differential operator. A key property is that the
zero-order term of this
operator is the transpose of the matrix $A$ defining the 1-form $\theta^a$.
\begin{lemma}
The Fr\'echet derivative $Z'$ of the vector field $Z$ may be uniquely
represented as the difference
\begin{equation}
\label{Zdecomp}
Z' = -A^\mathrm{t} + P\cdot R
\end{equation}
of the transpose of the matrix $A$ and of a factorized differential
operator \mbox{$P\cdot R$},
taking into account all the higher-order terms appearing in $Z'$.
\end{lemma}
\begin{proof}
The identity (\ref{Zdecomp}) is nothing else that a disguised form of the
Lagrange identity
\begin{equation}\label{Zidentita}
(\alpha,Z_{u}^{'}\phi)-(\phi,{Z_{u}^{'}}^{\ast}\alpha) = \frac{\D}{\D
x}B(\alpha,\phi)
\end{equation}
used to define the formal adjoint of the operator $Z'$. In this identity
$\alpha$ and $\phi$ are
arbitrary, and the bracket denotes the usual scalar product in
$\Rdoppia^n$. We notice that, by \eqref{campoZ}
the vector \mbox{${Z_{u}^{'}}^{\ast}(e_l)$} is the opposite of the Euler
operator associated
with the lagrangian
density $h^l$,
\begin{displaymath}
{Z_{u}^{'}}^{\ast}(e_l) = -\frac{\delta h^l}{\delta u}
\end{displaymath}
and we write the identity (\ref{Zidentita}), for \mbox{$\alpha=e_l$}, in
the operator form
\begin{displaymath}
-{h^l}^{'}(\phi) + \sum_{b=1}^n\phi^b\frac{\delta h^l}{\delta u^b} =
{\frac{\D }{\D x}}B(e_l,\phi)
\end{displaymath}
where ${h^l}^{'}$ is the Fr\'echet derivative of the scalar differential
operator $h^l$. One
easily recognizes in this equation the identity (\ref{Zdecomp}) by recalling
that
\begin{displaymath}
A^l_b = \frac{\delta h^l}{\delta u^b}.
\end{displaymath}
\end{proof}
The identity (\ref{Zdecomp}) allows to perform the fourth step in our
program rather easily. By
using once the operator form of the Lie derivative of $P$ we obtain
\begin{eqnarray*}
L_Z(P) & = & - Z^\prime{\Punto} P-P{\Punto}Z^{\prime\ast} \\
& = & A^\mathrm{t}{\Punto}P+P{\Punto}A + P{\Punto}(R^\ast-R){\Punto}P.
\end{eqnarray*}
 \\

\noindent\textbf{Steps 5 and 6:} \emph{the horizontal bivector $Q_H$ and
the $2$-form
$\omega$.}\\
By subtracting this identity from the basic representation formula
(\ref{Qnuovissimo}), we obtain
\begin{eqnarray}
Q & = & L_Z(P) + P{\Punto}(\Gamma+R-R^\ast){\Punto}P \\ \nonumber
& = & L_Z(P) + P{\Punto}\Omega{\Punto} P.\label{unarapp}
\end{eqnarray}
It allows to identify the 2-form $\omega$ with the restriction to the
symplectic leaves of $P$
of the differential operator \mbox{$-\Omega$}, where
\begin{displaymath}
\Omega = \Gamma+R-R^\ast
\end{displaymath}
defined on $M$. The explicit computation of this form is algorithmic, as
shown by the examples
given in the next section. We can thus conclude that we have a systematic
procedure to identify the
2-form $\omega$.\\

\noindent\textbf{Step 7:} \emph{the $2$-form $\omega$ is exact.}\\
The last steps are now performed along a well-established path. The
closure of the 2-form
$\Omega$ follows from the cocycle condition \mbox{$[P,Q]=0$}. By using the
operators form of this
condition we obtain:
\begin{eqnarray*}
[P,Q](\alpha,\beta,\gamma) & = & [P,P{\Punto}\Omega{\Punto}
P](\alpha,\beta,\gamma) \\
& = & \langle\alpha,P{\Punto}\Omega_u'(P\beta;P\gamma)\rangle +
\langle\beta,P{\Punto}\Omega_u'(P\gamma;P\alpha)\rangle +
\langle\gamma,P{\Punto}\Omega_u'(P\alpha;P\beta)\rangle \\
& = &\langle\alpha,
P{\Punto}[\Omega_u^\prime(P\beta;P\gamma) - \Omega_u^\prime(P\gamma;P\beta) +
{\Omega_u^{\prime}}^{\ast}(P\beta;P\gamma)]\rangle \\
& = & 0.
\end{eqnarray*}
Therefore $Q$ is a cocycle iff $\Omega$ verifies the equation
\begin{equation}\label{secondachiusura}
P{\Punto}[\Omega_u^\prime(P\beta;P\gamma) - \Omega_u^\prime(P\gamma;P\beta) +
{\Omega_u^{\prime}}^{\ast}(P\beta;P\gamma)]=0
\end{equation}
for any choice of the arguments $\beta$ and $\gamma$. Let us fix
$\beta$. We can regard
the previous equation as a differential equation on $\gamma$ of the form
\begin{displaymath}
({\frac{\D }{\D x}}{\Punto} T{\Punto} {\frac {\D }{\D x}}) \gamma = 0
\end{displaymath}
where $T$ is a suitable differential operator depending on $\beta$. By the
argument already used in
discussing equation (\ref{primachiusura}) we see that this equation can be
verified by any $\gamma$
only if \mbox{$T=0$}. This give rise to a new differential system on
$\beta$. Once again it can be
satisfied by any $\beta$ only if the equations are identically vanishing.
Thus we conclude that the
operator \eqref{secondachiusura} holds iff
\begin{displaymath}
\Omega_u^\prime(\phi;\psi) - \Omega_u^\prime(\psi;\phi) +
{\Omega_u^{\prime}}^{\ast}(\phi;\psi) = 0
\end{displaymath}
for any choice of the arguments $\phi$ and $\psi$. This is the
\emph{closure} condition for the
2-form $\Omega$.\\

\noindent\textbf{Step 8:} \emph{the potential $\theta$.}\\
Since we are working on a manifold with simple topology, by the Poincar\'e
lemma we
can affirm the existence of a 1-form $\theta$ such that \mbox{$\omega = \D
\theta$}. In our
particular context the 1-form $\theta$ may be represented as a
vector-valued differential operator
\begin{displaymath}
\theta = \theta(u,u_x,\ldots)
\end{displaymath}
and its exactness, in operator formalism, may be explicitly written as
\begin{equation}\label{secondaesattezza}
\Omega = {\theta_u^{\prime}}^{\ast} - \theta_u^\prime
\end{equation}
where $\theta_u^\prime$ is, as usual, the Fr\'echet derivative of the
1-form $\theta$. Like in the
finite-dimensional case, the operator $\theta$ can be reconstructed from
$\Omega$ by a quadrature.
The formula
\begin{equation}\label{latheta}
\theta = -\int_0^1\Omega_{\lambda u}(\lambda u) \D\lambda
\end{equation}
means that we must apply the matrix differential operator $\Omega$,
evaluated at the point
\mbox{$\lambda u$} on the vector \mbox{$\lambda u$} itself. Then we must
integrate, terms by
terms, the resulting vector-valued differential operator
\mbox{$\Omega_{\lambda u}(\lambda u)$},
depending on $\lambda$, on the interval \mbox{$[0,1]$}. Applications of
this formula will be given
in the next section.\\

We have finally achieved our goal. By inserting the representation
(\ref{secondaesattezza}) of the
2-form $\Omega$ in the representation formula (\ref{unarapp}) of the
cocycle $Q$ we obtain
\begin{eqnarray*}
Q & = & L_Z(P) + P{\Punto}({\theta_u^{\prime}}^{\ast} -
\theta_u^\prime){\Punto}P \\
& = & L_Z(P) + L_{P\theta}(P) \\
& = & L_{Z+P\theta}(P)
\end{eqnarray*}
showing that the cocycle $Q$ is a coboundary. Furthermore we obtain the
explicit formula
\begin{equation}\label{questocampo}
X=Z+P\theta
\end{equation}
for the potential $X$ of $Q$, as in the finite-dimensional case. The
proposition stated at the
beginning of this section is thus completely proved.

To prepare the discussion on Dubrovin's conjecture, to be performed in
the final section, it
remains to understand what relation connects the class of homogeneous
bivectors considered by
Dubrovin (and described in \secref{sec:1}) to the class of bivectors
considered in
this section.

\begin{lemma}
The class of Dubrovin's cocycles is strictly contained in the present class of
cocycles.
\end{lemma}
\begin{proof}
The point is to show that the homogeneity assumption (together with the
cocycle condition) entails
the involutivity condition (\ref{cond1}) used to define our class of
cocycles. To prove this
result we exploit the well-known property that, for every cocycle, the bracket
\mbox{$\{C^a,C^b\}_Q$} is still a Casimir function of $P$. This means that
\begin{displaymath}
P \D (\{C^a,C^b\}_Q) = 0
\end{displaymath}
that is
\begin{displaymath}
{\frac{\D }{\D x}}\frac{\delta E^{ab}}{\delta u^l} = 0.
\end{displaymath}
Let us write
\begin{equation}
\frac{\delta E^{ab}}{\delta u^l}=A^{ab}_l.
\end{equation}
By the above condition the functions $A^{ab}_l$ are constant, and therefore
\begin{displaymath}
E^{ab} = A^{ab}_l u^l + {\frac{\D }{\D x}} K^{ab}.
\end{displaymath}
Accordingly
\begin{displaymath}
\{C^a,C^b\}_Q = \int_0^1 A^{ab}_l u^l \D x .
\end{displaymath}
The homogeneity condition of Dubrovin entails \mbox{$A^{ab}_l=0$}, since
$E^{ab}$ should have at least degree one. So the combined action of the
cocycle and of
the homogeneity condition entails the involutivity (\ref{cond1}), as
required.
\end{proof}

We finally notice that the vector field $Z$ and the $1$-form $\theta$
associated with the homogeneous cocycle $Q$ are themselves homogeneous
operators, due to \eqref{hamiltoniane} and \eqref{latheta}. Thus we can end
our discussion by stating the following proposition
\begin{proposition}
All cocycles in Dubrovin's class are coboundaries, and their potentials are
homogeneous operators.
\end{proposition}
In our opinion, this is the deep reason for the validity of Dubrovin's
conjecture.

\section{Three examples}\label{sec:4}

As first example we consider again the homogeneous third-order scalar
differential operators
(\ref{pidue}). We have already shown that they are coboundaries by
guessing the form of the
vector field $X$, inside the class of homogeneous vector fields. Presently
we want to rediscover
systematically this vector field, by using the previous algorithm. We
recall the main steps of
this approach.

\begin{enumerate}
\item The starting point is the representation formula $A^\mathrm{t} {\Punto}
P + P {\Punto} A + P
{\Punto} \Gamma {\Punto} P.$ It allows to identify the matrix $A$ and the
2-form
$\Gamma$.
\item The next step is to exploit the pieces of information encoded into
the matrix $A$.
Its columns are exact
1-forms, and their potentials are the Hamiltonians with Lagrangian
densities $h^a$. They allow to
define the transversal
vector field \mbox{$Z=-h^a Z_a$}.
\item The Fr\'echet derivative $Z'$ of this vector field is the last object
to be
analyzed. Through the
second representation formula \mbox{$Z'=-A^\mathrm{t} + P {\Punto} R$}, it
allows to identify the
operator $R$, which defines the deformation \mbox{$\Omega = \Gamma + R -
R^\ast$} of $\Gamma$ we
are interested in.
\item At this point we compute the potential $\theta$ of the 2-form
$\Omega$. The
vector field $X$ is
given by the formula \mbox{$X=-H^a Z_a + P\theta$}.
\end{enumerate}

For the example at hands, the first representation formula reads
\begin{eqnarray*}
P_2 &=& [2\alpha(u)u_{xx}+\alpha^{'}(u){u_x}^2]{\frac{\D}{\D x}} +
{\frac{\D}{\D x}}\Punto[2\alpha(u)u_{xx}+\alpha^{'}(u){u_x}^2]\\
&& + {\frac{\D}{\D x}}\Punto[\beta(u){\frac{\D}{\D x}}
+ {\frac{\D}{\D x}}\Punto\beta(u)]\Punto{\frac{\D}{\D x}}
\end{eqnarray*}
where \mbox{$\alpha(u)$} and \mbox{$\beta(u)$} are related to the previous
coefficients \mbox{$E(u)$} and \mbox{$G(u)$} according to
\begin{eqnarray*}
\alpha(u) &=& \frac{1}{4}(G(u)-\frac{1}{2}E^{'}(u)) \\
\beta(u) &=& \frac{1}{2}E(u).
\end{eqnarray*}
Since \mbox{$P=\frac{\D}{\D x}$}, the ``matrix'' $A$ is simply the scalar
function
\begin{displaymath}
A(u;u_x,u_{xx}) = 2\alpha(u)u_{xx}+\alpha^{'}(u){u_x}^2 .
\end{displaymath}
We recognize in this expression the Euler operator associate with the
Lagrangian density
\begin{displaymath}
h(u;u_x) = -\alpha(u){u_x}^2 .
\end{displaymath}
Consequently the vector field $Z$ is given by
\begin{displaymath}
Z(u;u_x) = \alpha(u){u_x}^2 = \frac{1}{4}(G(u)-\frac{1}{2}E^{'}(u)){u_x}^2 .
\end{displaymath}
Its  Fr\'echet derivatives is
\begin{displaymath}
Z_u' = -2\alpha(u)u_{xx}-\alpha^{'}(u){u_x}^2+2\frac{\D}{\D x}[\alpha(u)u_x]
\end{displaymath}
and therefore
\begin{eqnarray*}
Z_u' + A^\mathrm{t} & = & 2\frac{\D}{\D x}[\alpha(u)u_x] .
\end{eqnarray*}
In this way we obtain the operator \mbox{$R=2\alpha(u)u_x$}. Since
\mbox{$R^\ast=R$}, the 2-form
$\Omega$ is simply given by
\begin{displaymath}
\Omega = \beta(u)\frac{\D}{\D x} + \frac{\D}{\D x}\Punto\beta(u).
\end{displaymath}
This form is exact and its potential is given by
\begin{displaymath}
\theta(u;u_x) = -\beta(u)u_x = -\frac{1}{2}E(u)u_x .
\end{displaymath}
It can be computed either by direct inspection or by using the equation
\begin{eqnarray*}
\theta & = & -\int_0^1 \Omega_{\lambda u}(\lambda u) d\lambda \\
& = & -\int_0^1 [\beta(\lambda u){\frac{\D}{\D x}}(\lambda u) +
\frac{\D}{\D x}(\beta(\lambda u))]d\lambda.
\end{eqnarray*}
Finally, the vector field $X$ is given by
\begin{displaymath}
X = Z + P\theta = -\frac{1}{2}E(u)u_{xx} +
(\frac{1}{4}G(u)-\frac{5}{8}E^{'}(u)){u_x}^2.
\end{displaymath}
It coincides with the vector field already obtained at the beginning of
\secref{sec:3}.\\

As a second example let us consider the non-homogeneous third-order
scalar differential operator
\begin{equation}
Q={\frac{\D ^3}{\D x^3}}+2u{\frac{\D }{\D x}}+u_x.
\end{equation}
It is the well-known second Hamiltonian operator of the KdV hierarchy. By
writing $Q$ in the form
$$Q=u\cdot {\frac{\D }{\D x}}+{\frac{\D }{\D x}} \Punto u+{\frac{\D
}{\D x}}\left({\frac{\D }{\D x}}\right)\Punto {\frac{\D }{\D x}}$$
we identify
$$A(u)=u$$
$$\Gamma ={\frac{\D }{\D x}}.$$
Once again, we recognize in $A$ the Euler operator associate with the
lagrangian density
$$h(u)=\frac{1}{2}u^2 .$$
Consequently
$$Z=-\frac{1}{2}u^2 .$$
Its Fr\'echet derivative verifies the equation
$$Z^{'}+A^{t}=-u+u=0 .$$
Hence $R=0$, and $\Omega =\frac{\D}{\D x}$. The potential $\theta$ of
this $2$-form is
$$\theta =-\frac{1}{2}u_x.$$
The potential $X$ associated with the operator $Q$ is then
\begin{equation}
X=-\frac{1}{2}u^2-\frac{1}{2}u_{xx}.
\end{equation}
\\

The final example concerns a non-homogeneous matrix-valued bivector $Q$. We
consider the pair of Poisson bivectors
\begin{equation}
P = \left(
\begin{array}{cc}
0 & 1 \\
1 & 0
\end{array}
\right) \frac{\D}{\D x}
\end{equation}
and
\begin{equation}\label{QBoussineq}
Q = \left(
\begin{array}{cc}
-\frac{\D ^3}{\D x^3}+2u\frac{\D}{\D x}+u_x  &
3v\frac{\D }{\D x}+2v_x  \\
\\
3v\frac{\D}{\D x}+v_x  &
\begin{array}{c}
\frac{\D ^5}{\D x^5} -\frac{10}{3}u\frac{\D ^3}{\D x^3}
-5u_{x}{\frac{\D ^2}{\D x^2}}\\
+(\frac{16}{3}u^2 -3u_{xx})\frac{\D}{\D x} +
(\frac{16}{3}uu_{x}-{\frac{2}{3}}u_{xxx})
\end{array}
\end{array}
\right)
\end{equation}
associated with the Boussineq hierarchy. It is well-known that $P$ is a
coboundary of
$Q$. Indeed $P$ is the Lie
derivative of $Q$ along the vector field
\begin{displaymath}
X_{-1} : \left\{
\begin{array}{c}
\dot{u} = 0 \\
\dot{v} = \frac{1}{2}
\end{array}
\right .
\end{displaymath}

Our aim is presently to show that $Q$ is a coboundary of $P$,
and we want to compute
explicitly its potential $X_{1}$:
\begin{displaymath}
Q= L_{X_{1}}(P).
\end{displaymath}
The boring part is the identification of the matrix $A$ and of the
operator $\Gamma$.
We start from the representation formula
\begin{equation}\label{rappBoussineq}
\begin{array}{rcl}
Q & = & E + S{\Punto}\frac{\D }{\D x} + \frac{\D }{\D x}{\Punto} S + \\
&& \frac{\D }{\D x}{\Punto} [\Lambda_0 + (\Lambda_1 {\Punto}\frac{\D }{\D
x}+\frac{\D
}{\D x}{\Punto} \Lambda_1) +
(\Lambda_2{\Punto} \frac{\D ^2}{\D x^2}+\frac{\D ^2}{\D x^2}{\Punto}
\Lambda_2)\\
 &&+(\Lambda_3{\Punto} \frac{\D ^3}{\D x^3}+\frac{\D ^3}{\D x^3}{\Punto}
\Lambda_3)]{\Punto}\frac{\D }{\D x}
\end{array}
\end{equation}
valid for any fifth-order bivector. By comparison with \eqref{QBoussineq},
we obtain
\begin{displaymath}
\begin{array}{ccc}
E = \left( \begin{array}{cc}
0 & \frac{1}{2}v_x \\
-\frac{1}{2}v_x & 0
\end{array} \right) \;
&
S = \left( \begin{array}{ccc}
u && \frac{3}{2}v \\
\frac{3}{2}v && \frac{8}{3}u^2-\frac{2}{3}u_{xx}
\end{array} \right) \;
&
\Lambda_0 = 0
\\
\\
\Lambda_1 = \left( \begin{array}{cc}
-\frac{1}{2} & 0 \\
0 & -\frac{5}{3}u
\end{array} \right)
&
\Lambda_3 = \left( \begin{array}{ccc}
0 && 0 \\
0 && \frac{1}{2}
\end{array} \right)
&
\Lambda_2 = 0
\end{array}
\end{displaymath}
We notice that $E$ is a total derivative with respect to $x$, guaranteeing that
\mbox{$\{C^a,C^b\}_Q=0$}. Since
\begin{equation}
E =\frac{\D}{\D x}\Punto
\left(
\begin{array}{cc}
0 & \frac{1}{2}v \\
-\frac{1}{2}v & 0
\end{array}
\right)+
\left(
\begin{array}{cc}
0 & -\frac{1}{2}v \\
\frac{1}{2}v & 0
\end{array}
\right)\Punto
\frac{\D}{\D x}
\end{equation}
we obtain the first representation formula of the operator $Q$
\begin{eqnarray*}
Q &=&\frac{\D}{\D x}\Punto
\left(
\begin{array}{ccc}
u && 2v \\
v &&  \frac{8}{3}u^2-\frac{2}{3}u_{xx}
\end{array}
\right)+
\left(
\begin{array}{ccc}
u && v \\
2v &&  \frac{8}{3}u^2-\frac{2}{3}u_{xx}
\end{array}
\right)
\Punto
\frac{\D}{\D x} \\
&&+{\frac{\D}{\D x}}{\Punto}
\left( \begin{array}{cc}
-\frac{\D}{\D x} & 0 \\
0 & \frac{\D^3}{\D^3 x}-\frac{10}{3}u\frac{\D}{\D x}-\frac{5}{3}u_x
\end{array} \right)
\Punto \frac{\D}{\D x} .
\end{eqnarray*}
Replacing the operator $\frac{\D}{\D x}$ by the bivector $P$, we
obtain the second representation formula
\begin{eqnarray*}
Q &=& P\cdot
\left(
\begin{array}{ccc}
v && \frac{8}{3}u^2-\frac{2}{3}u_{xx} \\
u && 2v
\end{array}
\right)+
\left(
\begin{array}{ccc}
v && u \\
\frac{8}{3}u^2-\frac{2}{3}u_{xx} && 2v
\end{array}
\right)
\Punto P \\
&&+P{\Punto}
\left( \begin{array}{cc}
\frac{\D^3}{\D^3 x}-\frac{10}{3}u\frac{\D}{\D x}-\frac{5}{3}u_x & 0 \\
0 & -\frac{\D}{\D x}
\end{array} \right)
\Punto P .
\end{eqnarray*}
It follows that
\begin{displaymath}
A = \left(
\begin{array}{ccc}
v && \frac{8}{3}u^2-\frac{2}{3}u_{xx} \\
u && 2v
\end{array}
\right)
\end{displaymath}
and
\begin{displaymath}
\Gamma = \left( \begin{array}{cc}
\frac{\D^3}{\D^3 x}-\frac{10}{3}u\frac{\D}{\D x}-\frac{5}{3}u_x & 0 \\
0 & -\frac{\D}{\D x}
\end{array} \right) .
\end{displaymath}
At this point we just repeat the usual scheme. We notice that the entries
of the columns of $A$
are the Euler operators associated with the Lagrangian densities
\begin{eqnarray*}
h^1(u,v) & = & uv \\
h^2(u,v;u_x,v_x) & = & v^2 + \frac{8}{9}u^3 + \frac{1}{3}u_x^2
\end{eqnarray*}
respectively. Then we use the transversal symmetries
\begin{eqnarray*}
z_1: && \left\{ \begin{array}{c}
\dot{u} = 1 \\
\dot{v} = 0
\end{array} \right.\\
z_2: && \left\{ \begin{array}{c}
\dot{u} = 0 \\
\dot{v} = 1
\end{array} \right.
\end{eqnarray*}
to build the vector field
\begin{displaymath}
Z:\; \left\{ \begin{array}{rcl}
\dot{u} &=& -uv \\
\dot{v} &=& -v^2 - \frac{8}{9}u^3 - \frac{1}{3}u_x^2.
\end{array} \right.
\end{displaymath}
Its Fr\'echet derivative verifies the equation
\begin{eqnarray*}
Z' + A^\mathrm{t} & = &
\left( \begin{array}{ccc}
-v && -u \\
-\frac{8}{3}u^2-\frac{2}{3}u_x\frac{\D}{\D x} && -2v
\end{array} \right) +
\left( \begin{array}{ccc}
v && u \\
\frac{8}{3}u^2-\frac{2}{3}u_{xx} && 2v
\end{array} \right) \\
& = &
\frac{\D}{\D x} \Punto \left( \begin{array}{cc}
0 & 0 \\
-\frac{2}{3}u_x & 0
\end{array} \right).
\end{eqnarray*}
So the operator $R$ is given by
\begin{displaymath}
R = \left(
\begin{array}{cc}
-\frac{2}{3}u_x & 0 \\
0 & 0
\end{array} \right).
\end{displaymath}
Since \mbox{$R=R^\ast$} we finally get \mbox{$\Omega=\Gamma$}. Its
potential $\theta$ is:
\begin{displaymath}
\theta = \left(
\begin{array}{c}
\frac{5}{3}uu_x - \frac{1}{2} u_{xxx} \\
\frac{1}{2} v_x
\end{array}
\right).
\end{displaymath}
Consequently
\begin{displaymath}
X_1:\; \left\{
\begin{array}{rcl}
\dot{u} &=& -uv + \frac{1}{2} v_{xx} \\
\\
\dot {v} &=& -v^2 - \frac{8}{9}u^3 + \frac{4}{3}u_x^2 + \frac{5}{3}uu_{xx} -
\frac{1}{2}u_{xxxx}.
\end{array}
\right.
\end{displaymath}

Let us finally consider the third vector field
\begin{displaymath}
2X_0 = [X_{-1},X_1]
\end{displaymath}
It is a conformal symmetry of both Poisson bivectors $P$ and $Q$. Indeed
\begin{eqnarray*}
L_{X_0}(P) &=& \frac{1}{2}P   \\
L_{X_0}(Q) &=& \frac{1}{2}Q.
\end{eqnarray*}
Furthermore, the vector fields \mbox{$(X_{-1},X_0,X_1)$} satisfy the
commutation relations
\begin{displaymath}
[X_{-1},X_0]=X_1 \qquad [X_1,X_0]=-X_{-1}.
\end{displaymath}
Therefore by the present algorithm we have constructed the
\mbox{$\mathfrak{sl}(2)$}-subalgebra of the $\mathcal{W}$-algebra
associated with the Boussineq
hierarchy. This remark suggests that the method used in this paper are
potentially very useful in analyzing and classifying Poisson pencils on
bihamiltonian manifolds.

\section{Proof of Dubrovin's conjecture}\label{sec:5}
The key idea for proving the conjecture is to reduce the Jacobi identity
\mbox{$[P_\epsilon,P_\epsilon]=0$} to a sequence of
cohomological equations. This is possible on a manifold of hydrodynamic
type due to the results of
\secref{sec:3}. The outcome is a peculiar representation of the
coefficients of the deformation $P_\epsilon$ in terms of vector fields.

\begin{proposition}
A sequence of homogeneous vector fields $X_k$ may be associated with every
homogeneous deformations
$P_\epsilon$, in such a way that the coefficients $P_k$ of the Taylor
expansion of $P_\epsilon$ are written as iterated derivatives of the given
bivector $P_0$. To this
end consider the Lie derivatives associated with the vector fields, and
construct with them the
operator
\begin{displaymath}
T_k = \sum_{j_1 + 2j_2 + \cdots + kj_k = k}
\frac{L^{j_k}_{X_k}}{J_k!}\cdots\frac{L^{j_1}_{X_1}}{J_1!}
\end{displaymath}
to be referred to as the Schur polynomial of order $k$ associated with the
given sequence of vector
fields. Then
\begin{equation}\label{ktermine}
P_k = T_k(P_0).
\end{equation}
\end{proposition}
\begin{proof}
Let us first check the formula for \mbox{$k=1$}. We know that the first
coefficient $P_1$ is an
homogeneous bivector verifying the cocycle condition \mbox{$[P_1,P_0]=0$}.
Hence, by the final
proposition of \secref{sec:3} there exist an homogeneous vector field
$X_1$, such that
\mbox{$P_1=L_{X_1}(P_0)$}. This proves the first case of identity
(\ref{ktermine}).

To prove by induction the remaining cases , we use the identity
\begin{equation}\label{trasforma}
T_k([P,P]) = \sum^k_{\begin{array}{c}
\scriptstyle{j,l = 0} \\
\scriptstyle{j+l=k}
\end{array}}
[T_j(P),T_l(P)] .
\end{equation}
It follows from the transformation law
\begin{displaymath}
\psi_{\epsilon\ast}([P,P]) = [\psi_{\epsilon\ast}(P),\psi_{\epsilon\ast}(P)]
\end{displaymath}
with respect to the special one parameter family of local diffeomorphisms
\begin{displaymath}
\phi_\epsilon^{(k)} : M \rightarrow M
\end{displaymath}
constructed as follows. First we compose the flows
\mbox{$(\phi_{t_1},\ldots,\phi_{t_k})$}
associated with the vector fields \mbox{$(X_1,\ldots,X_k)$} so to obtain
the multiparameter family
of local diffeomorphisms
\begin{equation}\label{multipar}
\phi^{(k)}_{t_1,\cdots,t_k} = \phi_{t_k}\circ \cdots \circ \phi_{t_1} .
\end{equation}
Then we reduce this family by setting
\begin{equation}
t_j = \epsilon^j .
\end{equation}
By expanding \eqref{multipar} in powers of $\epsilon$, and by equating the
coefficients of
$\epsilon^k$ we obtain exactly \eqref{trasforma}.

Assume presently that the representation (\ref{ktermine}) is true for the
first $n$ coefficients
\mbox{$(P_1,\ldots,P_n)$}. To prove that it is also true for $P_{n+1}$ we
consider
\eqref{trasforma} for \mbox{$k=n+1$}. We notice that this equation holds
for any choice of the
vector fields \mbox{$(X_1,\ldots,X_{n+1})$}. In particular it holds also
for \mbox{$X_{n+1}=0$}. Let
us denote by
\begin{displaymath}
\hat{T}_{n+1} = T_{n+1}|_{X_{n+1}=0}
\end{displaymath}
the restriction of the operator $T_{n+1}$ to the first $n$ vector fields of
the sequence. Then we
can write
\begin{equation}\label{trasforma cappello}
\hat{T}_{n+1}([P_0,P_0]) = \sum^{n+1}_{\begin{array}{c}
\scriptstyle{j,l = 0} \\
\scriptstyle{j+l={n+1}}
\end{array}}
[\hat{T}_j(P_0),\hat{T}_l(P_0)] .
\end{equation}
By assumption \mbox{$[P_0,P_0]=0$}, and
\begin{displaymath}
P_l = T_l(P_0) = \hat{T}_l(P_0) \qquad \forall l=1,\ldots,n .
\end{displaymath}
Therefore \eqref{trasforma cappello} becomes:
\begin{displaymath}
2[P_0,\hat{T}_{n+1}(P_0)] +\sum^n_{\begin{array}{c}
\scriptstyle{j,l = 1} \\
\scriptstyle{j+l=n+1}
\end{array}}
[P_j,P_l] = 0.
\end{displaymath}
Let us compare this equation with
\begin{displaymath}
2[P_0,P_{n+1}] +\sum^n_{\begin{array}{c}
\scriptstyle{j,l = 1} \\
\scriptstyle{j+l=n+1}
\end{array}}
[P_j,P_l] = 0
\end{displaymath}
expressing the Jacobi identity \mbox{$[P_\epsilon,P_\epsilon]=0$} at the
order \mbox{$n+1$} in
$\epsilon$. It takes the form of a cocycle condition:
\begin{displaymath}
[P_0,P_{n+1}-\hat{T}_{n+1}(P_0)]=0 .
\end{displaymath}
Therefore there exist a vector field $X_{n+1}$ such that
\begin{displaymath}
P_{n+1} = L_{X_{n+1}}(P_0) + \hat{T}_{n+1}(P_0) = T_{n+1}(P_0) .
\end{displaymath}
By induction this proves the representation formula (\ref{ktermine}) for
any $k$.
\end{proof}

To end the proof of Dubrovin's conjecture it is sufficient now to notice
that the infinite
sequence of identities
\begin{equation}\label{ktermine2}
P_k=T_k(P_0)
\end{equation}
means that
\begin{equation}\label{diffeomorfi}
P_\epsilon = \phi_{\epsilon\ast}(P_0)
\end{equation}
for the limit $\phi_\epsilon$ of the sequence of local diffeomorphism
$\phi_\epsilon^{(k)}$ for
\mbox{$k\rightarrow\infty$}. Indeed, according to the theory of ``Lie
transform'',
\eqref{ktermine2} are nothing else than the Taylor expansion of
\eqref{diffeomorfi} in powers
of $\epsilon$. We have then obtained a constructive proof of Dubrovin's
conjecture. The relation
\begin{equation}
\phi_\epsilon^{(k)} =
\phi_{\epsilon^k}^{[X_k]}\circ\cdots\circ\phi_\epsilon^{[X_1]}
\end{equation}
gives the approximation, at order $k$ of the trivializing map
\mbox{$\phi_\epsilon:M\rightarrow
M$} we were looking for.

\section*{Acknowledgments}
We sincerely thank B. Dubrovin for introducing us to the problem of
deformation of Poisson
manifolds of hydrodynamic type. We also thank G. Falqui for many useful
discussion. We finally
thank the Istituto Nazionale di Alta Matematica of Rome, who supported a
meeting on the geometry of Frobenius
manifolds, giving us the occasion to meet all together and discuss the problem.

\end{document}